\documentclass[conference,a4paper]{IEEEtran}

\ifCLASSINFOpdf
\else
\fi
\usepackage[left=1.57cm,right=1.57cm,top=0.95cm,bottom=2.54cm]{geometry}
\usepackage{listings}
\usepackage{setspace}
\usepackage{graphicx}
\usepackage{tabularx,booktabs}
\usepackage{dingbat}
\usepackage{diagbox}
\usepackage{multirow} 
\usepackage[hyphens]{url}
\usepackage[hidelinks,breaklinks]{hyperref}
\usepackage{xcolor}
\usepackage{amsfonts, amsmath, amsthm, amssymb}
\usepackage{subcaption}
\hypersetup{breaklinks=true}
\usepackage{tikz}
\usepackage{textcomp}
\usepackage{lipsum}
\usepackage{svg}
\IEEEoverridecommandlockouts
\newcommand\copyrighttext{%
  \footnotesize \textcopyright 2026 IEEE.  Personal use of this material is permitted.  Permission from IEEE must be obtained for all other uses, in any current or future media, including reprinting/republishing this material for advertising or promotional purposes, creating new collective works, for resale or redistribution to servers or lists, or reuse of any copyrighted component of this work in other works.}
\newcommand\copyrightnotice{%
\begin{tikzpicture}[remember picture,overlay]
\node[anchor=south,yshift=10pt] at (current page.south) {\fbox{\parbox{\dimexpr\textwidth-\fboxsep-\fboxrule\relax}{\copyrighttext}}};
\end{tikzpicture}%
}
\usepackage{graphicx}
\graphicspath{{figures/}}
\begin{document}

%
% paper title
% Titles are generally capitalized except for words such as a, an, and, as,
% at, but, by, for, in, nor, of, on, or, the, to and up, which are usually
% not capitalized unless they are the first or last word of the title.
% Linebreaks \\ can be used within to get better formatting as desired.
% Do not put math or special symbols in the title.
\title{Transformer-Based MCS Prediction for 5G Multicast-Broadcast Services (MBS)}

% author names and affiliations
% use a multiple column layout for up to three different
% affiliations
\author{
    \IEEEauthorblockN{Kasidis Arunruangsirilert, Jiro Katto}
    \IEEEauthorblockA{Department of Computer Science and Communications Engineering, Waseda University, Tokyo, Japan
    \\\{kasidis, katto\}@katto.comm.waseda.ac.jp}
}
% conference papers do not typically use \thanks and this command
% is locked out in conference mode. If really needed, such as for
% the acknowledgment of grants, issue a \IEEEoverridecommandlockouts
% after \documentclass
% for over three affiliations, or if they all won't fit within the width
% of the page, use this alternative format:
% 
%\author{\IEEEauthorblockN{Michael Shell\IEEEauthorrefmark{1},
%Homer Simpson\IEEEauthorrefmark{2},
%James Kirk\IEEEauthorrefmark{3}, 
%Montgomery Scott\IEEEauthorrefmark{3} and
%Eldon Tyrell\IEEEauthorrefmark{4}}
%\IEEEauthorblockA{\IEEEauthorrefmark{1}School of Electrical and Computer Engineering\\
%Georgia Institute of Technology,
%Atlanta, Georgia 30332--0250\\ Email: see http://www.michaelshell.org/contact.html}
%\IEEEauthorblockA{\IEEEauthorrefmark{2}Twentieth Century Fox, Springfield, USA\\
%Email: homer@thesimpsons.com}
%\IEEEauthorblockA{\IEEEauthorrefmark{3}Starfleet Academy, San Francisco, California 96678-2391\\
%Telephone: (800) 555--1212, Fax: (888) 555--1212}
%\IEEEauthorblockA{\IEEEauthorrefmark{4}Tyrell Inc., 123 Replicant Street, Los Angeles, California 90210--4321}}
% use for special paper notices
%\IEEEspecialpapernotice{(Invited Paper)}
% make the title area

\maketitle

\copyrightnotice
% As a general rule, do not put math, special symbols or citations
% in the abstract
\setstretch{0.95}
\begin{abstract}
The deployment of 5G Multicast-Broadcast Services (MBS) is emerging as a critical technology for spectral-efficient UHD content delivery and serving as a promising solution to modernize CATV deployment. However, unlike unicast networks that rely on RLC-AM with HARQ retransmissions, MBS broadcast operates in RLC Unacknowledged Mode (RLC-UM), where the absence of a feedback loop means packet loss is permanent and immediately impacts user QoE. Conventional link adaptation algorithms, designed for unicast, typically aggressively maximize throughput and fail in this risk-intolerant environment, resulting in severe video stalls and rebuffering. To address this, we propose a lightweight Transformer-based framework that predicts the success probability of all 28 MCS indices over an upcoming video segment horizon. Utilizing a unique commercial network dataset with 0.5 ms slot-level granularity, we train our model using a custom Asymmetric Safety Loss function that penalizes channel overestimation to prioritize link stability. Experimental results show that our approach achieves a reliability score of 86.89\%, significantly outperforming standard AI baselines optimized for raw throughput (31.65\%) while maintaining a safe conservative bias. Furthermore, the model is optimized for real-time applications, demonstrating an inference time of less than 0.07 ms on COTS 5G-era smartphones.

\end{abstract}


% no keywords

\setstretch{0.907}

% For peer review papers, you can put extra information on the cover
% page as needed:
% \ifCLASSOPTIONpeerreview
% \begin{center} \bfseries EDICS Category: 3-BBND \end{center}
% \fi
%
% For peerreview papers, this IEEEtran command inserts a page break and
% creates the second title. It will be ignored for other modes.
\IEEEpeerreviewmaketitle

\vspace{-1.5mm}
\section{Introduction}

The advent of 5G New Radio (NR) has fundamentally transformed the telecommunications landscape. With its high throughput and low air-interface latency, 5G serves as a key enabler for a wide array of emerging use cases, ranging from industrial automation and the Internet of Things (IoT) to self-driving vehicles and telemedicine \cite{9906062}. Beyond facilitating these advanced mobility and novel applications, 5G also addresses foundational infrastructure challenges. Specifically, the modernization of last-mile connectivity presents a significant challenge for the telecommunications industry, particularly within legacy Multi-Dwelling Units (MDUs), such as apartments and condominiums, where structural constraints render the retrofitting of Fiber-to-the-Home (FTTH) cabling cost-prohibitive or physically infeasible. Consequently, Private 5G Fixed Wireless Access (FWA) has emerged as a critical alternative for bridging the digital divide.

%The modernization of last-mile connectivity presents a significant challenge for the telecommunications industry, particularly within legacy Multi-Dwelling Units (MDUs), such as apartments and condominiums, where structural constraints render the retrofitting of Fiber-to-the-Home (FTTH) cabling cost-prohibitive or physically infeasible. Consequently,  Private 5G Fixed Wireless Access (FWA) has emerged as a critical alternative for bridging the digital divide.

In this context, Cable Television (CATV) services act as reliability-first infrastructure, providing essential real-time information and national broadcasts that cannot be fully replaced by internet-based solutions \cite{10874994}. Hence, the industry is increasingly adopting 5G Multicast-Broadcast Services (MBS), which allow for the spectrally efficient transmission of Ultra-High Definition (UHD) content to multiple users via a 5G Radio Access Network (RAN) deployed for broadcast delivery \cite{9806729, 9950805, 10913885}. However, the architecture of 5G MBS in broadcast mode introduces new constraints absent in unicast communications. Unlike conventional commercial 5G NR networks, which are unicast networks utilizing the Acknowledged Mode of Radio Link Control (RLC-AM) with Hybrid Automatic Repeat Request (HARQ) enabled, MBS broadcast streams rely on RLC Unacknowledged Mode (RLC-UM) \cite{3GPP_38-331, 9772755}. This creates a strict pipe condition: the system lacks a feedback loop for retransmission at the MAC layer. While the system has the advantage of achieving a consistent and minimal End-to-End (E2E) latency, the failure mode manifests as packet loss in two different ways. 

\begin{figure*}[t!]
\centering\includesvg[width=0.9\linewidth,inkscapelatex=false]{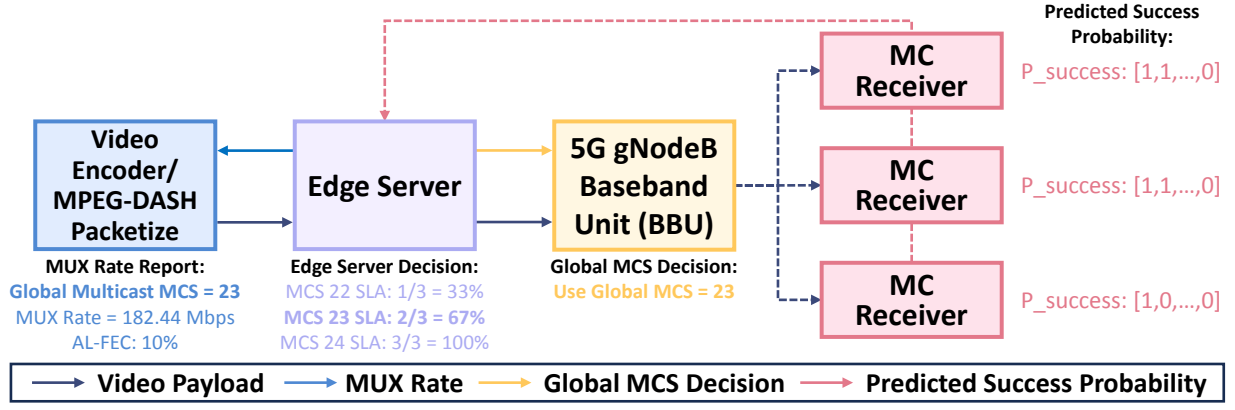}
  \setlength{\belowcaptionskip}{-2pt}
  \caption{System Architecture Overview (Multicast Part)}
  \label{fig:SystemArchitecture}
  \vspace{-6mm}
\end{figure*}

First, if the chosen Modulation and Coding Scheme (MCS) is too aggressive for the instantaneous channel quality, the receiver fails to decode the Physical Downlink Shared Channel (PDSCH), leading to data loss if the error rate exceeds Application Layer FEC (AL-FEC) limits. Second, in a CATV-replacement deployment scenario, MBS is operated on a dedicated network, in which the Common Frequency Resource (CFR) is fully allocated to broadcast without unicast contention. With a standard fixed MIMO Rank of 1 \cite{10913885}, the data rate is strictly proportional to the MCS. Therefore, if the incoming video bitrate exceeds this MCS-defined capacity, the gNodeB buffer saturates, forcing packet drops prior to transmission \cite{3GPP_38-323}. In MPEG-DASH, either scenario results in the corruption of a Group of Pictures (GOP), causing playback stalls and introducing user-perceived latency. Therefore, latency in MBS broadcast does not manifest as a function of queue depth, which typically increases E2E latency, but rather as a binary state defined strictly by packet loss (success or failure). While this failure state can be mitigated by switching to a more robust MCS, such an adjustment inherently reduces the overall data rate, resulting in degraded visual quality, as the application-layer video bitrate must be proportionally scaled down to match the constrained throughput of the air interface. \looseness=-1

Conventional 5G link adaptation algorithms are designed for unicast scenarios where HARQ masks prediction errors, allowing schedulers to maximize throughput aggressively. Applying these algorithms to the risk-intolerant, no-HARQ environment of MBS broadcast will result in severely degraded user's Quality of Experience (QoE). Therefore, reliable MBS operation requires a predictive mechanism that prioritizes link stability over raw throughput maximization. The system must forecast channel degradation with sufficient time to adjust the video encoder bitrate and MCS proactively. 

To realize this proactive adaptation, we envision a holistic cross-layer architecture where each Multicast (MC) Receiver predicts its own success probability distribution for all MCS indices and transmits this telemetry back to an Edge Server. As for future implementation, the Edge Server will aggregate these reports and dictate the global MCS decision to the gNodeB to satisfy the configured Multicast Service Level Agreement (SLA). From this selected MCS, the corresponding multiplex (MUX) data rate can be derived and communicated directly to the video encoder (see Fig. \ref{fig:SystemArchitecture}). Furthermore, our proposed method is intended to be used within an Adaptive MBS model \cite{10913885}, where users whose predicted channel qualities fall below the selected global MCS will be dynamically served by a supplementary univast stream. Because evaluating this complete end-to-end system (see Fig. \ref{fig:SystemArchitecture}) exceeds the scope of this paper, this work focuses strictly on the foundational component: the User Equipment (UE)-side prediction framework. To address these challenges, this paper presents the following core contributions:

\begin{enumerate}
  \item \textbf{Probabilistic Cross-Layer Forecasting:} We propose a novel, lightweight framework that predicts the full success probability distribution of all 28 MCS indices (MCS Table 2 \cite{3GPP_38-214}) over a future video GOP horizon, calibrated to compensate for signal propagation and system processing delays, enabling risk-aware real-time bitrate adaptation.
  \item \textbf{Micro-Granular Real-World Validation:} In contrast to previous studies, our approach is based on a unique commercial network dataset with 0.5 ms slot-level granularity. This allows us to capture ultra-fast channel dynamics often abstracted away in theoretical models.
  \item \textbf{Practical \& Standard-Compliant Design:} Our method relies exclusively on passive metrics measurable at the User Equipment (UE). It is designed for low-latency applications, achieving an inference time of $<0.07$ ms on Commercial Off-The-Shelf (COTS) smartphones to satisfy real-time constraints.
\end{enumerate}

\section{Background and Related Work}

\subsection{5G Multicast-Broadcast Services (MBS) Constraints}

3GPP Release 17 standardized 5G MBS to enable efficient point-to-multipoint delivery \cite{li_swanson_bampis_krasula_aaron_2020, stare_munier_van_2022, 9772755}.  Building upon this standard, Ito et al. previously proposed a hybrid MIMO architecture for Private 5G broadcasting \cite{10913885}, where users experiencing weak reception are dynamically switched to pre-allocated time slots for unicast transmission, which allows the gNodeB to maintain a more spectrally efficient MCS for the majority of users. However, this architecture does not solve the fundamental problem of estimating the correct MCS in the first place. Accurate prediction is crucial to determine which MCS is suitable for each user's specific channel quality. Therefore, without this precise estimation, the switching logic cannot function effectively. Because MBS Broadcast relies on RLC-UM without HARQ retransmissions, overestimating the channel quality causes the receiver to fail in decoding the payload. Consequently, packets are immediately discarded, leading to video stalls. Thus, a robust and risk-averse prediction method is necessary.

\subsection{Limitations of Existing MCS Prediction Methods}

Existing literature on 5G link adaptation largely focuses on unicast scenarios (RLC-AM), which are not applicable in the RLC-UM context. A significant portion of the research focuses on predicting intermediate channel metrics such as Channel Quality Indicator (CQI) or Signal-to-Noise Ratio (SNR), rather than the final packet success rate. For instance, Abdulhasan et al. proposed a Feed Forward Neural Network technique for CQI prediction \cite{7238169}, while Saija et al. utilized Machine Learning (ML) to forecast SNR values to assist in MCS selection \cite{9118097}. While these approaches improve the granularity of Channel State Information (CSI), they remain indirect and do not inherently predict the final packet success probability (BLER) across different MCS indices, nor do they account for the non-linear relationship between SNR and successful decoding in fading channels.

On the other hand, many works utilize ML to predict throughput or MCS directly, yet these models typically produce deterministic outputs. Minovski et al. and Batool et al. developed ML models using lower-layer metrics to predict downlink throughput \cite{9495144, 10577702}. Likewise, our previous work utilized a similar approach for uplink throughput prediction \cite{10849914}. Tsipi et al. evaluated algorithms like Random Forest and SVM for MCS prediction \cite{repec:spr:telsys:v:86:y:2024:i:4:d:10.1007_s11235-024-01158-x}. While achieving high accuracy in their corresponding contexts, these models output a single value without providing a risk profile or probability distribution. In an RLC-AM unicast environment, a deterministic prediction that slightly overestimates capacity is corrected by HARQ retransmissions. However, in the RLC-UM environment of MBS, such overestimation leads to immediate packet loss. Moreover, many of these models are trained on coarse-grained data, often aggregated at 1-second intervals or derived from simulations \cite{9495144, 10849914, 10.1145/3339825.3394938, 8361404}. This aggregated view is blind to microscopic dynamics, failing to capture the millisecond-level fast fading and physical channel coherence characteristics of 5G Sub-6 Bands (Frequency Range 1 or FR1). \looseness=-1

Finally, several studies propose system-level scheduling frameworks rather than predictive engines. Pocovi et al. addressed joint link adaptation for Ultra-Reliable Low-Latency Communications (URLLC) \cite{8361404}, while Ohseki et al. proposed fast outer-loop link adaptation (OLLA) schemes. While relevant to low-latency constraints, standard URLLC implementations often rely on redundancy mechanisms, such as packet duplication and URLLC-specific MCS Table (MCS Table 3 \cite{3GPP_38-214}). Additionally, OLLA mechanisms are inherently reactive, adjusting offsets based on previous ACK/NACK feedback. In MBS broadcast, the absence of per-user HARQ feedback renders conventional OLLA inapplicable. Consequently, these frameworks are not predictive models in themselves. They require an underlying, proactive mechanism to function in real-world environments where the channel relationship is unknown and dynamic.

\subsection{RLC-AM vs. RLC-UM Mismatch}

The common key limitation across the surveyed literature is the inherent design focus on unicast-based commercial 5G networks (RLC-AM) \cite{9148457, 10147378}. However, prediction frameworks for RLC-UM remain largely underexplored. In RLC-AM, failure manifests as increased latency due to retransmissions. In 5G MBS (RLC-UM), the failure mode is packet loss. Existing systems optimized for RLC-AM maximize spectral efficiency by tolerating a certain block error rate of 10\%, assuming HARQ will recover the error exceeding the threshold. Applying these systems to MBS results in a 10\% packet loss rate, which has to be corrected by AL-FEC, and any error exceeding the capability of AL-FEC will render high-bitrate video streams unwatchable. This fundamental mismatch necessitates a novel, probabilistic approach that penalizes overestimation to guarantee transmission success without retransmission mechanisms.

\section{Proposed Method}

\subsection{Data Collection}

As real-world data from a commercial 5G network is necessary for the training, testing, and validation of the proposed method, we utilized a Samsung Galaxy S22 Ultra 5G (SC-52C), equipped with the Snapdragon X65 5G Modem-RF System. However, unlike the target 5G MBS, which has a fixed MIMO rank of 1, commercial networks typically employ a dynamic MIMO rank (up to 4) to maximize the data rate. Since reconfiguration of commercial base stations was not possible, we modified the UE modem firmware by rewriting Non-Volatile (NV) items to disable three of the four receive antennas at the hardware level. This method physically powers down the corresponding Low Noise Amplifiers (LNAs) and receive paths, yielding an electrically equivalent native single-antenna (1Rx) device with identical fast-fading statistics, channel coherence, and absence of spatial diversity gain. This effectively forces a 4 × 1 MIMO configuration.

To quantify the performance impact of using a single receive antenna, preliminary experiments were conducted by utilizing the phone's internal \textit{ServiceMode} menu, which physically rewrites the corresponding NV items to the EFS partition, to switch between \textit{RX0 Only} mode, which uses only the primary receive antenna, and \textit{NR FORCED 4RX MODE}, which activates all four antennas, under \textit{RF TEST} submenu, enabling a direct comparison between two configurations. The UE was attached to a Huawei 64T64R Massive MIMO Active Antenna Unit (AAU) installed on a gNodeB located 800 meters away. For each antenna configuration, a two-minute HTTP GET test was performed using the \textit{Network Signal Guru (NSG)}, a professional 5G drive test tool, to commence downlink stress tests from an AIS Ookla Speedtest Server. This entire procedure was conducted at six different locations to cover both Line-of-Sight (LOS) and Non-Line-of-Sight (NLOS) scenarios, during off-peak hours (9:00 PM) to avoid interference and variations from network congestion. We ensured that the phone remained in the exact same position when switching between the two antenna configurations at each spot. The collected logs were then imported into \textit{AirScreen} software for analysis, where the average Channel State Information Signal-to-Interference-plus-Noise Ratio (CSI-SINR) was calculated for each configuration to determine the performance difference. The results revealed an average SINR penalty of \textbf{2.64 dB} when comparing the single-antenna configuration to the baseline four-antenna setup (see Table \ref{tab:RxCompare}). \looseness=-2

\begin{table}[!tbp]
\setstretch{0.75}
\caption{1 Rx vs 4 Rx CSI-SINR Comparison (dB)}
\vspace{-1.5mm}
\centering
\label{tab:RxCompare}
\resizebox{8cm}{!}{\begin{tabular}{@{}lccccccc@{}}
\toprule
Config./Test Spot & A & B & C & D & E & F & Avg.\\\midrule
4 Rx (Baseline) & 9.18 & 12.58 & 4.73 & 20.82 & 8.56 & 5.46 & 10.22\\
1 Rx & 5.04 & 10.54 & 1.13 & 19.25 & 6.08 & 3.48 & 7.58 \\\midrule
\textbf{Difference} & \textbf{4.14} & \textbf{2.04} & \textbf{3.60} & \textbf{1.58} & \textbf{2.48} & \textbf{1.98} & \textbf{2.64}\\
\bottomrule
\end{tabular}}
\vspace{-6mm}
\end{table}

\begin{figure}[t!]
\centering\includegraphics[width=0.85\linewidth]{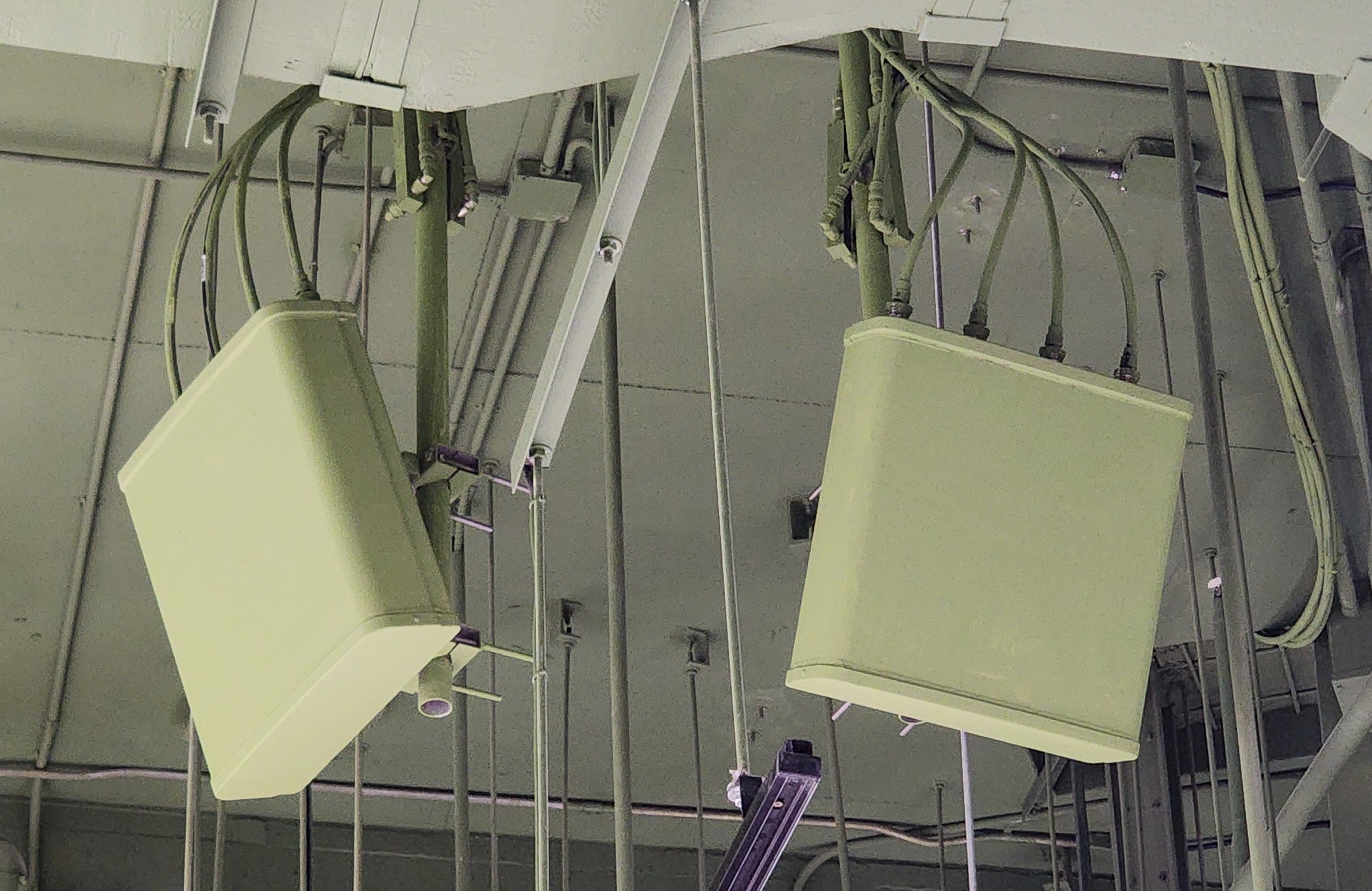}
  \setlength{\belowcaptionskip}{-12pt}

  \caption{AIS 4T4R Passive Antenna at Dusit Central Park}
  \label{fig:passiveAntenna}
\end{figure}

\begin{figure}[t!]
\centering\includegraphics[width=0.6\linewidth]{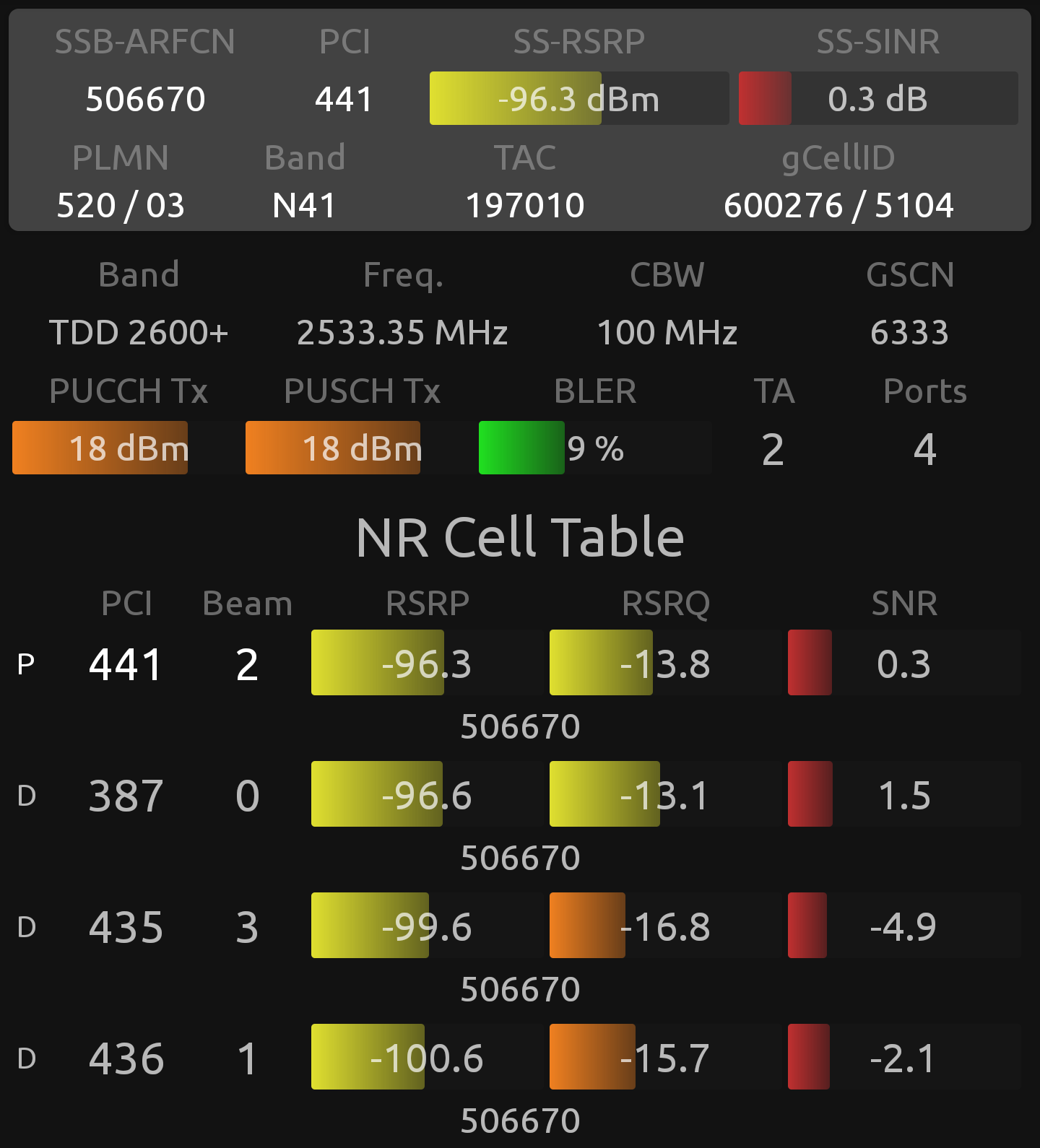}
  \setlength{\belowcaptionskip}{-14pt}

  \caption{NSG confirming the 4T4R passive panel antenna deployment topology at the data collection site (CSI-Port = 4). The NR Cell Table reveals multiple independently operated RRU, each identified by a distinct Physical Cell Identity such as PCI 441, 387, 435, and 436.}
  \label{fig:CellList}
  \vspace{-3mm}
\end{figure}
\begin{figure*}[t!]
\centering\includesvg[width=0.9\linewidth,inkscapelatex=false]{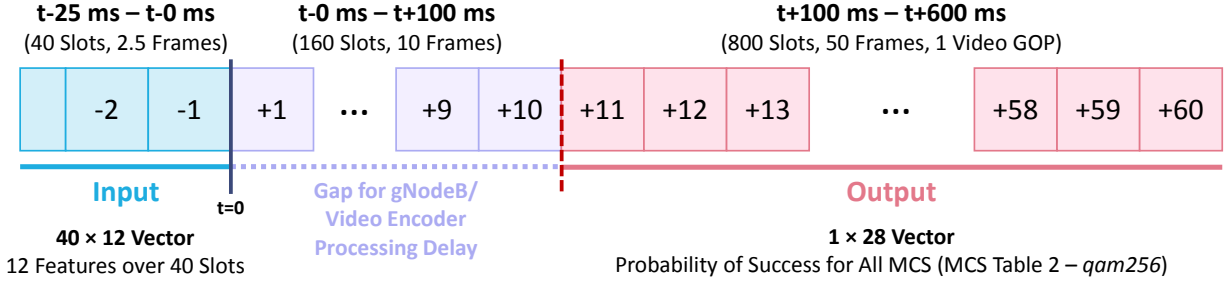}
  \setlength{\belowcaptionskip}{-2pt}
  \caption{Example temporal configuration of the proposed framework based on evaluated parameters. The model receives a $40 \times 12$ feature matrix extracted from a 25 ms observation window ($t_{obs}$, blue; $t-25$ ms to $t-0$). A 100 ms propagation and processing gap ($t_{delay}$, purple) explicitly accounts for gNodeB and video encoder propagation and processing delays. The final prediction is a $1 \times 28$ vector defining the success probability distribution for each Modulation and Coding Scheme (MCS) index across the 500 ms Video GOP horizon ($t_{GOP}$, pink; $t+100$ ms to $t+600$ ms).}
  \label{fig:Timeline}
  \vspace{-6mm}
\end{figure*}
With this performance baseline established, the primary data collection was conducted on the Advanced Info Service (AIS) 5G network in Thailand operating in the n41 band (2.5 GHz) with a 100 MHz bandwidth at Dusit Central Park, an indoor shopping mall. The mall officially opened on September 4, 2025. A preliminary site survey was conducted on September 7, 2025, to confirm the feasibility of the experiment, determine the network topology, and design the data collection procedure. However, unprecedented visitor volume during the opening week caused severe network congestion on both 4G LTE and 5G networks. Because MBS operates on dedicated resources without unicast contention, it was imperative to capture clean data where the maximum number of Resource Blocks (RBs) were scheduled to our test device. Therefore, to mitigate the impact of user load contention on spectral efficiency measurements, primary data collection was deferred until the initial visitor surge subsided. The actual experiments were conducted on September 16–17, 2025, by walking the premises during pre-opening hours (8:00-10:30 AM), yielding approximately five hours of clean data. \looseness=-1

Unlike most shopping malls, which utilize a Distributed Antenna System (DAS) or 2T2R Huawei Lampsite pRRU (Pico Remote Radio Units), the deployment in this mall consists of several independently operated 4T4R RRUs with passive panel antennas (see Fig. \ref{fig:passiveAntenna} and Fig. \ref{fig:CellList}), which mirrors the target MBS deployment, as opposed to the 64T64R Massive MIMO AAUs used for outdoor deployments. Consultations with an AIS network engineer confirmed the accuracy of our observed topology. Furthermore, the engineer clarified that this 4T4R setup is an interim deployment, expedited to guarantee coverage for the grand opening while a permanent, integrated infrastructure solution remains under development. We designed the walk path to ensure that the dataset covered all possible channel conditions, from Reference Signal Received Power (RSRP) -140 dBm to -44 dBm. 

Although the target application is RLC-UM, we performed data collection on the commercial RLC-AM network, which allowed us to capture HARQ feedback, specifically the Cyclic Redundancy Check (CRC) pass/fail states, which served as the ground truth for our predictive model. Similar to the preliminary experiments, we used the \textit{Network Signal Guru (NSG)} to execute a high-throughput HTTP GET stress test against the AIS Ookla SpeedTest server, ensuring the downlink channel remained saturated to capture the maximum MCS under varying channel conditions.

\subsection{Data Pre-Processing and Feature Engineering}

The raw logs captured by NSG were converted into Qualcomm's Diagnostic Log File (.dlf) format and processed using the \textit{Qualcomm Commercial Analysis Toolkit (QCAT)} to extract physical layer metrics at the slot level. A key challenge in processing telemetry data is the heterogeneity of reporting intervals. Some parameters such as Number of Resource Blocks (NumRB), MCS, CRC State, and CQI are reported every 0.5 ms via \textit{NR5G MAC PDSCH Status (0xB887)} and \textit{NR5G MAC CSF Report (0xB8A7)} packets, respectively. On the other hand, channel metrics like SINR and RSRP are reported at 20 ms and 160 ms intervals, via \textit{NR5G ML1 Searcher Measurement Database Update Ext (0xB97F)} and \textit{NR5G LL1 LOG SERVING SNR (0xB8D8)}, respectively. To align these metrics, we applied a Last-Observation-Carry-Forward (LOCF) interpolation to synchronize all features to the 0.5 ms master timeline. Additionally, data points associated with outdoor Physical Cell Identity (PCI) were filtered out to isolate the indoor environment. Moreover, the slots containing retransmitted data (MCS 28-31) were also filtered out. Notably, the network operates with a Time Division Duplex (TDD) frame structure of 7/2 (DL/UL) plus one special subframe for DL-UL switching, resulting in a 5 ms periodicity containing 8 downlink slots.

\begin{figure}[t!]
\centering\includegraphics[width=0.98\linewidth]{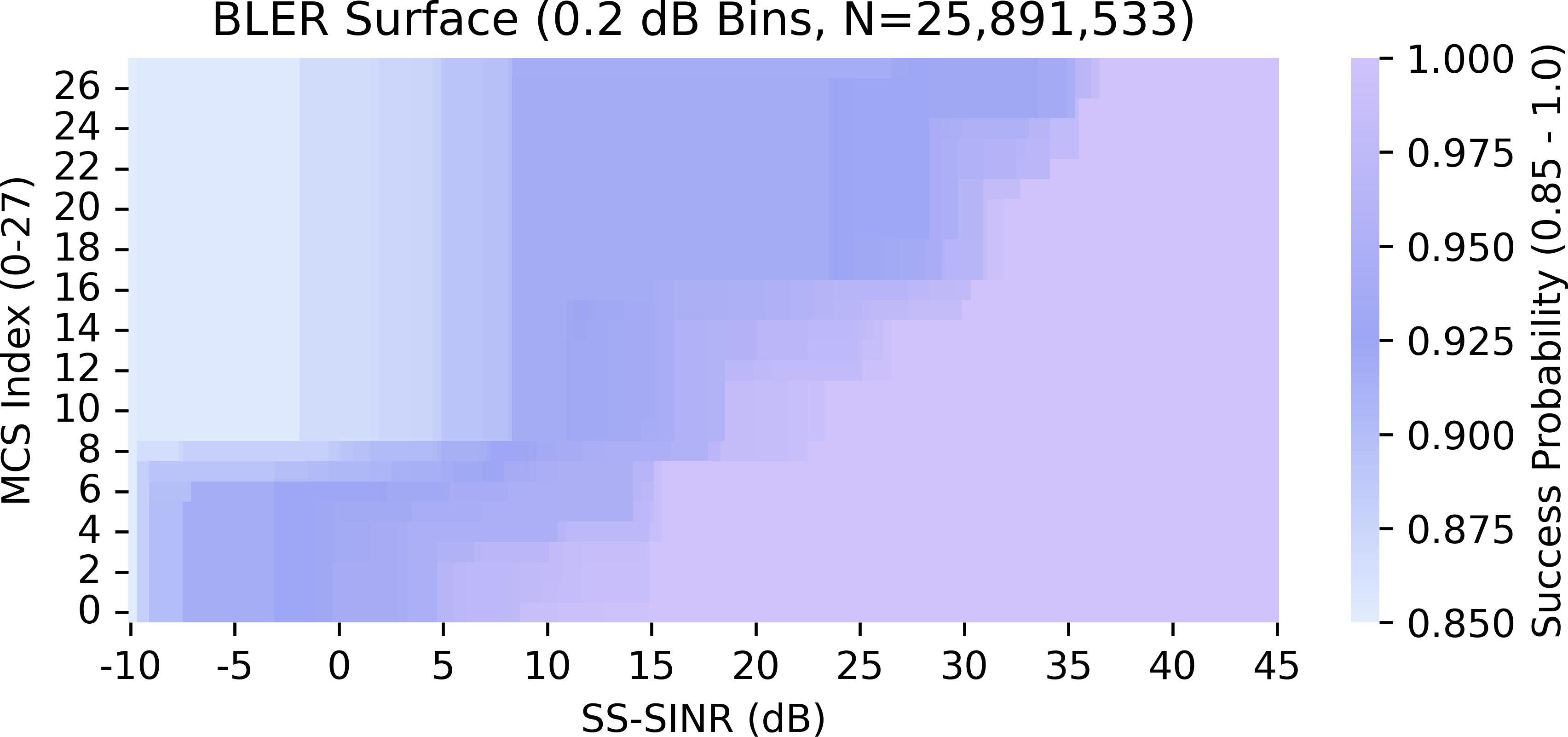}
  \setlength{\belowcaptionskip}{-14pt}
  \vspace{-5mm}
  \caption{MCS Index vs SS-SINR Heatmap}
  \label{fig:Heatmap}
  \vspace{-2mm}
\end{figure}

In our preliminary experiments conducted on a Massive MIMO base station, CSI-SINR was the primary metric used, as the AAU hardware is capable of forming dynamic, highly targeted RF beams on a per-user basis. When a UE is in \textit{RRC\_CONNECTED} state, the base station directs a narrow, user-specific beam toward the device, and the CSI Reference Signal (CSI-RS) is transmitted over the same beam to probe the channel condition that the PDSCH payload will experience. Hence, in this scenario, the CSI-SINR closely reflects the signal quality of the payload, revealing a strong correlation between CSI-SINR and the selected MCS indices. However, the 4T4R passive panel antenna deployment used in our actual data collection lacks a sufficient number of physical antenna elements required to perform dynamic, user-specific beamforming. Therefore, both the Synchronization Signal Block (SSB) and the PDSCH are transmitted over the same wide, static coverage beam. As a result, the SS-SINR, measured on the SSB, and the CSI-SINR, measured on the CSI-RS, experience similar propagation conditions.

As observed in Fig. \ref{fig:Heatmap}, this led to a stronger correlation between SS-SINR and the scheduled MCS index compared to CSI-SINR in this passive antenna environment. Therefore, we selected eight raw 5G parameters that are closely related: [NumRB, MCS, CRC\_State, SS\_RSRP, SS\_SINR, CSI\_RSRP, CSI\_SINR, DL\_CQI]. To capture temporal channel conditions and link stability, we engineered four additional features. These are categorized into state counters, which reset conditionally without a window limitation, and trend indicators, which are calculated over a rolling window of 16 slots (10 ms): 

\begin{itemize}
\item \textbf{Consecutive\_NACKs:} A cumulative counter of sequential packet failures. It increments on every CRC fail and resets to zero upon a successful transmission, serving as a direct metric of immediate link instability.
\item \textbf{Time\_Since\_Last\_NACK:} A cumulative counter of the number of slots since the last CRC failure. This metric quantifies the duration of the current stable state.
\item \textbf{MCS\_Trend:} The rolling average of the MCS index. This provides a smoothed baseline of the scheduler's recent behavior without instantaneous slot-to-slot jitter.
\item \textbf{CQI\_Trend:} The rolling average of the reported CQI, providing a smoothed representation of the recent channel quality baseline.
\end{itemize}

The objective of the model is to predict the probability of success for all valid MCS indices (0–27) over a future horizon, $t_{GOP}$, corresponding to the length of a video GOP. In this study, that window spans from $t+100$ ms to $t+600$ ms, matching the 500 ms GOP duration typically utilized in low-latency MPEG-DASH applications. An offset, $t_{delay}$, is incorporated to account for propagation and system processing delays. This ensures the gNB receives the prediction in time to make informed decisions on MCS selection and instruct the video encoder to adjust the bitrate accordingly. While a 100 ms delay is assumed for this study, this parameter is adjustable to suit specific application requirements. Finally, the observation window, $t_{obs}$, was set to 25 ms to balance historical context with model complexity.

Since the collected data contains only the result for the scheduled MCS at any given slot, we must infer the success probability for other indices. We apply a conservative monotonic logic. If a packet transmitted with MCS $m$ passes the CRC check, we assume that all lower, more robust MCS indices ($k<m$) would also have succeeded. Consequently, we increment both the trial and success counts for $m$ and all $k<m$. Conversely, if MCS $m$ fails, we do not assume that lower indices would have passed, as the gNodeB may have significantly overestimated the channel quality. In this case, we increment only the trial count for the specific MCS $m$, without inferring success for lower indices.

Let \(S_{k}\) be the success count and \(T_{k}\) be the total trial count for MCS index \(k\) within the prediction window. For each observed slot in the future horizon with scheduled MCS \(m_{obs}\) and CRC result \(c_{obs}\) (1 for Pass, 0 for Fail): 

\vspace{-5mm}
\begin{equation}
\text{If } c_{obs} = 0: \quad
\text{For } k = m_{obs}, \quad T_k \leftarrow T_k + 1
\end{equation}
\begin{equation}
\begin{split}
\text{If } c_{obs} = 1: \quad&
\forall k \leq m_{obs}, \\
&\begin{cases}
S_k \leftarrow S_k + 1 \
T_k \leftarrow T_k + 1
\end{cases}
\end{split}
\end{equation}

This logic mitigates the risk of introducing false positives into the dataset. Since a transmission failure at MCS $m$ does not guarantee success at $m-1$, as the actual channel quality might only permit a much lower MCS, we restrict data imputation strictly to confirmed successful transmissions. 

The temporal structure of the proposed framework is illustrated in Fig. \ref{fig:Timeline}. The input to the model is a sliding window of the last 40 slots ($t-25$ ms to $t-0$), resulting in an input array shape of $40 \times 12$. A 100 ms offset is incorporated to account for propagation and system processing delays. While this delay parameter is highly configurable to suit specific target applications, we deliberately selected 100 ms for this study. Our previous study indicates that typical round-trip ping times on commercial 5G networks range between 10 ms and 60 ms, averaging around 30 ms \cite{10118777}. A 100 ms gap comfortably absorbs this latency variance, ensuring the Edge Server receives the telemetry with sufficient lead time to make informed decisions on MCS selection and proactively instruct the video encoder to adjust its bitrate. Following this gap, the target label is a $1 \times 28$ vector representing the success probability \(P(Success)_{k}=S_{k}/T_{k}\) for each MCS index over the upcoming 500 ms Video GOP horizon. The final preprocessed dataset comprises 25.9 million data points. For model development, we partitioned this dataset into training (70\%), validation (15\%), and testing (15\%) subsets.

\subsection{Model Design and Training}

\begin{figure}[t!]
\centering\includesvg[width=0.98\linewidth,inkscapelatex=false]{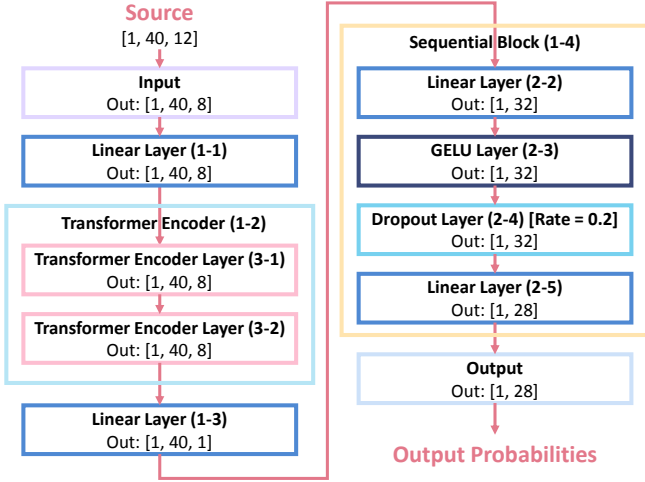}
  \setlength{\belowcaptionskip}{-14pt}
  \vspace{-5mm}
  \caption{Proposed Model Architecture}
  \label{fig:ModelArchitecture}
  \vspace{-2mm}
\end{figure}

To address the latency and reliability requirements of 5G MBS, we designed a lightweight Transformer-based model \cite{vaswani2023attentionneed}. The architecture is optimized for inference on resource-constrained UEs while effectively capturing the temporal characteristics of the physical channel (see Fig. \ref{fig:ModelArchitecture}). The model accepts an input tensor corresponding to $t_{obs}=25$ ms history (40 slots at 0.5 ms granularity) of the 12 input features. To account for propagation and processing delays, our framework incorporates a configurable 100 ms predictive offset. The input features are first linearly projected to a latent dimension of 8. To retain temporal order information, learnable positional embeddings are added to the projected input. The core of the network consists of two Transformer Encoder layers, utilizing Pre-Layer Normalization for training stability, a 2-head self-attention mechanism, and a feed-forward dimension of 32. Instead of relying on a single time-step, which can lose context from earlier channel variations, the model employs a pooling mechanism to aggregate the encoder outputs into a single context vector. Finally, this vector is passed through a Multi-Layer Perceptron (MLP) decoder with GELU activation to produce the $1 \times 28$ output vector representing the predicted success probability for each MCS index over the upcoming 500 ms Video GOP horizon. With a total of only 3,469 trainable parameters, the model is computationally efficient and suitable for on-device deployment. \looseness=-1 

A crucial innovation in our training process is the formulation of an Asymmetric Safety Loss ($L_{ASL}$) function tailored for the RLC-UM constraints of MBS, which penalizes channel overestimation to prioritize link stability. While the general principle of asymmetric penalization has recently proven effective for unicast URLLC link adaptation \cite{10364852}, our formulation applies this concept to a full multi-MCS probability distribution within a no-HARQ broadcast environment. Standard regression losses like Mean Squared Error (MSE) treat overestimation and underestimation symmetrically. However, in the RLC-UM context of MBS, overestimating the channel capacity results in immediate packet loss, whereas underestimating merely causes a minor reduction in spectral efficiency. The proposed loss function, which is based on MSE, can be defined as:
\vspace{3mm}

\centerline{\(L_{ASL} = \frac{1}{N} \sum_{i=1}^{N} w_i \cdot (\hat{y}_i - y_i)^2\)}
\vspace{2mm}

Where: 

\vspace{2mm}
\centerline{\(w_i =
\begin{cases}
    \lambda, & \text{if } (\hat{y}_i - y_i) > 0 \quad (\text{Overshoot}) \\
    1, & \text{otherwise}
\end{cases}\)}

\vspace{3mm}

Here, $\lambda$ is a penalty weight set to 1.4, which creates a gradient landscape that pushes the model to be conservative, effectively teaching it to prioritize link stability over raw throughput maximization. The model was trained using the AdamW optimizer with a weight decay of $1\times10^{-2}$ to prevent overfitting. We employed a OneCycleLR scheduler with a peak learning rate of $1\times10^{-3}$, a 30\% warmup phase, and cosine annealing. The training was conducted with a batch size of 512 for 10 epochs, each with 31,500 iterations, on an NVIDIA RTX PRO 6000 Blackwell Max-Q GPU. The checkpoint with the lowest validation loss was saved.

\subsection{Evaluation Method}

\subsubsection{Prediction Accuracy} \label{sec_PredictionAccuracyMethod}

%To rigorously assess the model's real-world performance, we developed a trace-driven simulator that emulates the behavior of 5G MBS scheduler, which operates on a 500 ms Group of Pictures (GOP) pacing. The simulation iterates through the validation dataset, which is the last 15\% of the data we collected, in steps of 800 slots, representing duration of one GOP. At 100 ms before the beginning of each GOP, a decision is made based on the data from the latest 40 slots, which accounted for the delay of the UE to informed the base station to make informed decision on MCS index, then enforce the rate control over the encoder for the upcoming GOP. This decision is then held constant and compared against the Ground Truth for the corresponding future 500 ms horizon starting at the beginning of GOP ($t+100$ to $t+600$ ms from the point of prediction). 

To rigorously assess the model's real-world performance, we developed a trace-driven simulator that emulates the behavior of a 5G MBS scheduler, which operates on a 500 ms GOP pacing. The simulation iterates through the testing dataset, which constitutes the final 15\% of the collected data, in steps of 800 slots, representing the duration of one GOP. At 100 ms before the beginning of each GOP, a decision is made based on the data from the latest 40 slots, accounting for the feedback latency required for the UE to report the predicted result, the base station to make an informed decision, and the video encoder to enforce rate control for the upcoming GOP. This decision is then held constant and compared against the ground truth for the corresponding future 500 ms horizon, which spans from $t+100$ ms to $t+600$ ms relative to the point of prediction.

While our model predicts the full success probability distribution for all 28 MCS indices, for the purpose of this paper, we evaluate performance by applying a decision policy with a 90\% success probability threshold. This threshold was selected because a BLER of 10\% is the target for link adaptation in enhanced Mobile Broadband (eMBB) use cases \cite{9732349}. However, a key advantage of our probabilistic framework is that this threshold is not fixed; it can be dynamically adjusted to meet diverse application requirements. For example, to match the reliability requirements in an Ultra-Reliable Low-Latency Communications (URLLC) scenario, the success threshold can be adjusted to reflect a target BLER of $10^{-5}$. \looseness=-1

We evaluate the performance of our model against four baselines to demonstrate its necessity and superiority. The first two represent simple engineering heuristics: a \textbf{Legacy Reactive Adaptation (LRA)} baseline uses the last observed MCS value from the input window, representing a standard non-predictive approach. The \textbf{Moving Average Window (MAW)} baseline uses the mean of the historical MCS values over the 40-slot window, acting as a simple smoothing heuristic. To ensure a fair comparison, the normalized input features were inverse-transformed to their original integer scale (0–27) for these two baselines, as they operate on a single real-world MCS value. 

To specifically isolate and validate the contributions of our probabilistic decision policy and asymmetric loss function, we introduce two additional AI-based baselines. The third is a \textbf{Deterministic AI} model, which uses the same Transformer architecture but is trained with standard MSE loss and configured to select the highest MCS with a probability of 50\% or higher, simulating a standard regression or classification model. The fourth, and most direct, comparison is an \textbf{MSE-Trained Transformer (MSE-T)}. This baseline uses the identical architecture and the same 90\% safety threshold as our proposed method, but similar to the previous case, was trained with the standard MSE loss function instead of our proposed asymmetric loss. This final baseline is designed to demonstrate that simply applying a powerful architecture is insufficient, and the risk-aware training objective is critical for achieving high reliability. \looseness=-2

The performance of each method is quantified using four key metrics. \textbf{RMSE (Tracking Accuracy)} measures the Root Mean Squared Error between the selected MCS and the ground truth. \textbf{Reliability Score} is the primary metric for MBS, defined as the percentage of GOPs where the selected MCS was less than or equal to the Ground Truth, thus avoiding packet loss. \textbf{Average Bias (Safety Margin)} calculates the mean difference between the selected MCS and the Ground Truth, where a negative value indicates a desirable conservative offset. Finally, \textbf{MAE (Average Error)} measures the average absolute deviation in MCS steps, providing an overall sense of prediction error magnitude.

\subsubsection{Inference Time}
\begin{table}[!tbp]
\setstretch{0.75}
\caption{Prediction Accuracy}
\vspace{-1.5mm}
\centering
\label{tab:accuracy}
\resizebox{7.5cm}{!}{\begin{tabular}{@{}lcccc@{}}
\toprule
\multirow{2}{*}{Method} & \multirow{2}{*}{RMSE} & Reliability & Average & \multirow{2}{*}{MAE}\\
&& Score & Bias & \\\midrule
\textbf{Proposed (Ours)} &3.1693&86.89\%&-1.8254&2.2978\\\midrule
LRA & 5.1401&83.69\%&-2.9341&3.5130\\
MAW& 4.7178&94.55\%&-3.4086&3.6002\\\midrule
Deterministic&2.8641&31.65\%&1.4555&2.1438\\
MSE-T&3.0110&83.81\%&-1.5312&2.1478\\
\bottomrule
\end{tabular}}
\vspace{-6mm}
\end{table}
To assess the feasibility of real-time deployment on resource-constrained UE, we evaluated the model's inference latency across a diverse range of COTS smartphones. The trained PyTorch model was converted to the TensorFlow Lite (TFLite) format to enable on-device execution. We used the AI Benchmark 6 to measure inference speed over 1,000 iterations per device, ensuring statistically stable results by mitigating operating system scheduling jitter. The test device pool spanned multiple hardware generations, including the Google Pixel 3, Huawei P40 Pro, Sony Xperia 1 III, Samsung Galaxy S22 Ultra, Samsung Galaxy Z Flip5, and Google Pixel 10 Pro XL (see Table \ref{tab:inference}). To maximize hardware acceleration, we utilized the Qualcomm QNN HTP (Hexagon Tensor Processor) backend for Snapdragon-based devices, while relying on the Android Neural Networks API (NNAPI) delegate for non-Qualcomm devices.

\section{Results and Analysis}

%	RMSE	Reliability	Bias	MAE
%Proposed				3.1693 86.89% -1.8254 2.2978
%LRA	5.1401	83.69%	-2.9341	3.513
%MAW	4.7178	94.55%	-3.4086	3.6002
%Deterministic	2.8641	31.65%	1.4555	2.1438
%MSE-T	3.011	83.81%	-1.5312	2.1478

\subsection{Prediction Accuracy}

The performance of our proposed model and the four baselines was evaluated using the trace-driven simulation described in Section \ref{sec_PredictionAccuracyMethod}. The results, summarized in Table \ref{tab:accuracy}, demonstrate the trade-offs between tracking accuracy and reliability in the context of 5G MBS. The heuristic-based baselines, LRA and MAW, serve as benchmarks for simple, non-AI approaches. While MAW achieves a high Reliability Score of 94.55\%, this safety comes at the cost of extremely poor tracking accuracy (RMSE of 4.7178) and a large conservative bias of -3.4086 MCS steps. This indicates that while averaging is safe, it is highly inefficient and slow to adapt to improving channel conditions, leading to significant underutilization of channel capacity. The LRA method performs poorly across all metrics, with a high RMSE and the lowest reliability among the safe methods, confirming its inadequacy for dynamic channel conditions.

The AI-based baselines highlight the importance of our specific design choices. The Deterministic model, which optimizes for the most likely outcome, achieves the lowest RMSE (2.8641), indicating excellent tracking precision. However, it fails catastrophically in reliability, with a score of only 31.65\%. Its large positive bias of +1.4555 confirms that it is dangerously aggressive, consistently overestimating channel capacity and causing packet loss in over two-thirds of all GOPs. This result proves that optimizing for traditional accuracy metrics is fundamentally unsuitable for the constraints of RLC-UM. The MSE-Trained Transformer (MSE-T) baseline, which uses our model architecture but is trained with a standard MSE loss function, isolates the impact of our asymmetric loss. While it achieves a low RMSE of 3.0110, its Reliability Score of 83.81\% is noticeably lower than our proposed method. In contrast, our proposed method, trained with the Asymmetric Safety Loss, strikes the most effective balance. It achieves a solid Reliability Score of 86.89\% while maintaining a competitive RMSE of 3.1693 and a low Mean Absolute Error (MAE) of 2.2978. The intentional negative bias of -1.8254 demonstrates that the model has successfully learned to be conservative, creating a necessary safety margin that directly translates into higher reliability and fewer video stalls. \looseness=-1

\subsection{Inference Time}

\begin{table}[!tbp]
\setstretch{0.8}
\caption{Inference Time on User Equipment}
\vspace{-1.5mm}
\centering
\label{tab:inference}
\resizebox{8.5cm}{!}{\begin{tabular}{@{}llcccc@{}}
\toprule
\multirow{2}{*}{Device} & \multirow{2}{*}{SoC} & Release&Avg. Infer & SD & \% of 5G \\
&&Year& Time (ms)& (ms)& TTI (0.5 ms)\\\midrule
Google Pixel 3&Snapdragon 845&2018&1.32&0.33&264             \\
Huawei P40 Pro&Kirin 990 5G&2020&0.07&0.07&14                \\
Sony Xperia 1 III&Snapdragon 888&2021&0.04&0.04&8            \\
Samsung Galaxy S22 Ultra&Snapdragon 8 Gen 1&2022&0.04&0.04&8 \\
Samsung Galaxy Z Flip5&Snapdragon 8 Gen 2&2023&0.04&0.04&8   \\
Google Pixel 10 Pro XL&Tensor G5&2025&0.03&0.03&6            \\
\bottomrule
\end{tabular}}
\vspace{-6.5mm}
\end{table}

To validate the model's suitability for real-time, on-device execution, we measured its inference latency on a range of commercial smartphones after converting the model to the TensorFlow Lite (TFLite) format. The results are detailed in Table \ref{tab:inference}. The performance across different hardware generations clearly demonstrates the feasibility of our approach on modern devices. The oldest device, a 2018 Google Pixel 3 (4G LTE UE) with a Snapdragon 845 SoC, required 1.32 ms for a single inference, consuming 264\% of a 0.5 ms 5G Transmission Time Interval (TTI). This confirms that devices without dedicated neural processing hardware are not suitable for this real-time application.

However, all subsequent devices performed exceptionally well. The Huawei P40 Pro (2020), utilizing the Android NNAPI, achieved an inference time of 0.07 ms (14\% of a TTI). Modern Snapdragon-powered devices, such as the Samsung Galaxy S22 Ultra and Z Flip5, leveraged the Qualcomm QNN HTP backend to achieve a consistent inference time of just 0.04 ms, or 8\% of a TTI. Moreover, the Google Pixel 10 Pro XL, the latest generation device, with its Tensor G5 SoC, demonstrated a further reduction to 0.03 ms (6\% of a TTI). These results confirm that our lightweight Transformer architecture is highly efficient and that inference latency is negligible on current and future generations of 5G UEs, leaving processing headroom for other device functions.

\section{Conclusion and Future Work}

This paper presented a Transformer-based MCS prediction framework specifically designed for the strict constraints of 5G Multicast-Broadcast Services (MBS). We identified a fundamental mismatch between conventional RLC-AM-optimized algorithms and the RLC-UM environment of MBS leads to unacceptable packet loss when applied to broadcast streams. To resolve this, we introduced a probabilistic prediction model trained with an Asymmetric Safety Loss ($L_{ASL}$) function, which effectively teaches the network to prioritize link reliability over raw throughput maximization.

Our trace-driven evaluation demonstrated that while standard deterministic AI models achieve high tracking accuracy, they fail catastrophically in reliability (31.65\%) due to aggressive overestimation. In contrast, our proposed method achieved a reliability score of 86.89\%, providing the necessary safety margin to minimize video stalls in live streaming applications. Additionally, we validated the practical feasibility of our approach on diverse COTS devices, confirming that modern smartphones can execute the model with negligible latency ($<8\%$ of a TTI).

In future work, we aim to expand the generalizability of our model beyond the specific indoor environment and hardware configuration used in this study. We plan to validate the model's performance across a wider variety of User Equipment (UE) models to ensure robustness against different modem chipsets and antenna designs. Furthermore, we will extend our data collection and testing to cover a broader range of deployment environments and mobility profiles, including stationary usage, outdoor walking, and high-mobility vehicular scenarios such as buses and trains, to capture the full spectrum of 5G channel dynamics and evaluate model robustness against varying Doppler shifts and complex outdoor multipath fading environments.

Most critically, this UE-side prediction model will serve as the foundation for the Adaptive MBS architecture. Future work will focus on implementing and evaluating this complete end-to-end system, featuring an Edge Server that aggregates these predictions to make SLA-driven global MCS decisions. This server will simultaneously provide dynamic multiplexing (MUX) rate feedback to the video encoder and orchestrate seamless unicast fallback for users experiencing channel conditions below the selected global MCS threshold. By integrating these components, we aim to realize a fully proactive, risk-aware Adaptive MBS solution.

\vspace{-0.5mm}
\section*{Acknowledgement}
\vspace{-0.5mm}
This paper is supported by the Ministry of Internal Affairs and Communications (MIC) Project for Efficient Frequency Utilization Toward Wireless IP Multicasting. Additionally, the authors would like to express their gratitude to \textbf{PEI Xiaohong} of \textit{Qtrun Technologies} for providing \textit{Network Signal Guru (NSG)} and \textit{AirScreen}, the cellular network drive test software used for result collection and analysis in this research.

% use section* for acknowledgment

% trigger a \newpage just before the given reference
% number - used to balance the columns on the last page
% adjust value as needed - may need to be readjusted if
% the document is modified later
%\IEEEtriggeratref{8}
% The "triggered" command can be changed if desired:
%\IEEEtriggercmd{\enlargethispage{-5in}}

% references section

% can use a bibliography generated by BibTeX as a .bbl file
% BibTeX documentation can be easily obtained at:
% http://mirror.ctan.org/biblio/bibtex/contrib/doc/
% The IEEEtran BibTeX style support page is at:
% http://www.michaelshell.org/tex/ieeetran/bibtex/
%\bibliographystyle{IEEEtran}
% argument is your BibTeX string definitions and bibliography database(s)
%\bibliography{IEEEabrv,../bib/paper}
%
% <OR> manually copy in the resultant .bbl file
% set second argument of \begin to the number of references
% (used to reserve space for the reference number labels box)
\setstretch{0.85}
\Urlmuskip=0mu plus 1mu\relax
\bibliographystyle{IEEEtran}
\bibliography{b_reference}

% that's all folks
\end{document}